\documentclass[aps,prl,superscriptaddress,twocolumn]{revtex4-2}

\usepackage{graphicx}
\usepackage{bm}
\usepackage{amssymb,amsfonts,amsmath}
\usepackage{hyperref}
\usepackage{float}

\makeatletter
\def\maketitle{
\@author@finish
\title@column\titleblock@produce
\suppressfloats[t]}
\makeatother

\begin{document}

\title{Quantum-geometry-enabled Landau--Zener tunneling in singular flat bands}
\author{Xuanyu Long}
\affiliation{Department of Materials Science and Engineering, University of Utah, Salt Lake City, UT 84112, USA}
\author{Feng Liu}
\email{ftiger.liu@utah.edu}
\affiliation{Department of Materials Science and Engineering, University of Utah, Salt Lake City, UT 84112, USA}
\date{\today}

\begin{abstract}
Flat-band materials have attracted substantial interest for their intriguing quantum geometric effects. Here we investigate how singular flat bands (SFBs) respond to a static, uniform electric field and whether they can support single-particle dc transport. By constructing a minimal two-band lattice model, we show that away from the singular band crossing point (BCP), the Wannier--Stark (WS) spectrum of the flat band is well captured by an intraband Berry phase $\Phi_{\mathrm{B}}$. The associated WS eigenstates are exponentially localized along the field direction, precluding dc transport. In contrast, near the BCP the interband Berry connection becomes prominent and drives Landau--Zener tunneling, which bends the flat-band WS ladder and delocalizes the SFB wavefunctions. Remarkably, this regime is governed solely by the maximal quantum distance $d$ through two geometric phases $(\theta,\varphi)$: $\theta$ characterizes the tunneling rate and $\varphi$ acts as a generalized Berry phase. These results highlight the essential role of quantum geometry in enabling nontrivial transport signatures in SFBs. 
\end{abstract}

\maketitle

\emph{Introduction.\textemdash} 
Quantum geometry plays an important role in determining the physical behavior of quantum materials, which is most prominent in flat band systems~\cite{natrev22,prl23qg,npj25qg}. At the first glance, a flat band with infinite effective mass cannot conduct, even when it is partially filled. However, in the presence of attractive electron-electron interaction, flat bands are shown to exhibit a finite superfluid weight~\cite{natrev22,prl23qg,npj25qg,nc15sc,prl16sc,prb17sc,prl20sc,prb22sc}. The underlying physics is the disruption of the perfect destructive interference by interaction~\cite{natrev22} or external field~\cite{25SFB}, which is described by the quantum geometry tensor~\cite{80qgt,10qgt}, whose real part is quantum metric and imaginary part is proportional to Berry curvature. Such disruption is also clearly exemplified by the anomalous Landau levels of singular flat bands (SFBs)~\cite{prb03ll,nature,sfbrev21,25SFB}, which is determined by the maximal Hilbert-Schmidt quantum distance $d$~\cite{pra96dis} that characterizes the singularity of wavefunctions around the band crossing point (BCP)~\cite{nature,sfbrev21}, and also independently by the real-space distance in some cases~\cite{25SFB}. 

On the other hand, for non-interacting particles, whether the flat band can have nonzero dc conductivity is still under debate~\cite{prb23fbcond,prb13fbcond,prb15cond,prb20fbcond,prb21fbcond,prb22fbcond}, especially for the case of SFB with BCP. One outstanding question is whether electric field can also disrupt the destructive interference of SFB, thereby enabling nontrivial single-particle transport?

The motion of a Bloch electron under a static, uniform electric field can be described by the Wannier--Stark (WS) spectrum~\cite{pr60ws,rmp62ws} with equidistant energy levels, which provides a useful framework for understanding transport phenomena represented by Bloch oscillation~\cite{Bloch29} and multiband generalizations~\cite{Zener34,njp06bz,prb12lz,wsspec,prl16bz,prb23bz,prl23bo,prb25bo,prb25ws,25fblz}. When the interband coupling is strong, for example near the Dirac point in graphene~\cite{prb12lz,wsspec}, Landau--Zener tunneling (LZT)~\cite{lz1,lz2} is pronounced, which is characterized by the tunneling possibility $P_{\mathrm{LZ}}$ and the Stokes phase $\varphi_{\mathrm{S}}$ accumulated in the non-adiabatic process~\cite{pr10lz,pra97sto,pra15lz}. The phase factor may result in quantum interference referred to as Landau--Zener--St\"uckelberg interferometry~\cite{pr10lz}, enabling ultrafast and reliable control of quantum systems~\cite{nat13lz,nat17lz,prl24lz}. The LZT in graphene provides a condensed-matter realization of the Schwinger mechanism~\cite{prd08sch,prb10sch,prb16sch}, where electron--hole pair production contributes to transport.

In general, LZT is related to non-Abelian Berry connections~\cite{wsspec,pra15lz,prl23bo,prb25ws}. A very recent work~\cite{25fblz} simulated Bloch oscillations in kagome and Lieb lattices, emphasizing the distinct LZT behavior near quadratic and linear BCPs. In this Letter, we instead investigate the WS states of a generic SFB model and demonstrate that the LZT is governed solely by $d$, and fully captured by two geometric phases. This in turn leads to a delocalization of the SFB wavefunctions, indicative of nontrivial flat-band transport~\cite{prb25bo}. 

Our calculations show that away from the BCP, the flat-band WS energies are governed by the Berry phase $\Phi_{\mathrm{B}}$, consistent with the modern theory of polarization~\cite{rmpberry,polar1,polar2,rmppolar}; the associated WS eigenstates are exponentially localized, forbidding dc transport. Near the BCP, two geometrical phases $(\theta,\varphi)$ emerge that fully characterize LZT: $\theta$ describes the tunneling rate and gap opening in the WS spectrum; $\varphi$ serves as a generalized Berry phase that describes the bending of the flat-band WS ladder. We further verify these results in a realistic SFB system, the kagome lattice.

\emph{Model construction of an SFB.\textemdash}
To study the WS spectrum and eigenstates of an SFB under an electric field $\bm{F}$, we first construct a minimal lattice model with tunable $d$. The effective continuum Hamiltonian for an isotropic SFB is~\cite{nature,25SFB}
\begin{align}
H_{\mathrm{SFB}} = 
t\begin{pmatrix}
  k_x^2 + (1-d^2) k_y^2 &\hspace{-0.5cm} d \sqrt{1-d^2} k_y^2 - i \xi d k_x k_y \\[0.5em]
  d \sqrt{1-d^2} k_y^2 + i \xi d k_x k_y &\hspace{-0.5cm} d^2 k_y^2
\end{pmatrix},
\label{eq:sfb}
\end{align}
where $k_x$ and $k_y$ are lattice momenta, and $\xi$ is the chirality of the flat-band wavefunction. For lattice realizations, we focus on the simplest case in which the Berry curvature vanishes everywhere except at the singular BCPs. Each BCP can be viewed as a Berry flux center, analogous to Dirac/Weyl points~\cite{prbFBwav,berryflux}. To ensure that the total Berry phase over the Brillouin zone vanishes, for arbitrary $d\in(0,1)$ the BCPs occur in pairs~\cite{prbFBwav} with opposite chirality. 

Following Ref.~\cite{fbcons}, we replace $k_x$ and $k_y$ in Eq.~\eqref{eq:sfb} by periodic functions to construct a shortest-range lattice Hamiltonian with tunable $d$:
\begin{widetext}
\begin{align}
H(\bm{k}) =
 t\left(\begin{array}{cc}
  (1-\cos 2 k_x)(1+\cos k_y)/4 + (1-d^2)(2-2\cos k_y) & d \sqrt{1-d^2} (2-2\cos k_y) - i d\, \sin k_x\, \sin k_y \\[0.5em]
  d \sqrt{1-d^2} (2-2\cos k_y) + i d\, \sin k_x\, \sin k_y & d^2 (2-2\cos k_y)
\end{array}\right).
\label{eq:sfbl}
\end{align}
\end{widetext}
The band structure [Fig.~\ref{fig1}(a)] contains an exactly flat band tangent to a dispersive band at two singular BCPs located at $\Gamma$ $(0,0)$ and X $(\pi,0)$. The flat band has zero energy, and the dispersive band is expressed as:
\begin{align}
E_c(\bm{k}) = t\,[9-7\cos k_y - \cos 2k_x (1+\cos k_y)]/4.
\label{eq:ec}
\end{align}
Expanding Eq.~\eqref{eq:sfbl} around $\Gamma$ and X reproduces Eq.~\eqref{eq:sfb} with opposite chirality $\xi=\pm 1$; $\xi$ thus plays the role of a valley index. We also compute the wavefunctions and Berry connections $\bm{A}_{mn} (\bm{k})=i\langle u_m (\bm{k}) |\frac{\partial}{\partial \bm{k}}| u_n (\bm{k}) \rangle$ for Eq.~\eqref{eq:sfbl} [see Sec.~I of the supplementary materials (SM)~\cite{sm}].

In real space, Eq.~\eqref{eq:sfbl} corresponds to a square lattice (lattice constant $a=1$) with two orbitals A and B per unit cell located on the same lattice site [Fig.~\ref{fig1}(b)]. The flat band supports compact localized states (CLSs), obtained by Fourier transforming the unnormalized $k$-space wavefunction (see Eq. S4 of the SM~\cite{sm}). These CLSs exhibit perfect destructive interference, with the hopping parameters encoded in Eq.~\eqref{eq:sfbl}.

\emph{Hamiltonian under an electric field.\textemdash}
We apply a uniform electric field $\bm{F}$ along the $y$-direction, so that the two valleys are not coupled by $\bm{F}$ and can be treated independently. Translational symmetry along $x$ is preserved, and $k_x$ remains a good quantum number. We therefore block diagonalize the Hamiltonian by $k_x$ and reduce the problem to an effective one-dimensional (1D) system (see Sec.~II A of the SM~\cite{sm}). For each $k_x$, we consider a finite 1D chain along $y$ with an onsite potential $F y_i$ proportional to the $y$-coordinate of the $i^{\rm th}$ orbital; the WS spectrum follows from diagonalizing this finite-size Hamiltonian.

Equivalently, Fourier transforming to $k$-space (Sec.~II B of the SM~\cite{sm}) gives the coupled eigenvalue equations (setting $e=1$):
\begin{align}
E\, \phi_0 &= i F \frac{\partial \phi_0}{\partial k_y} + F A^y_{00}\, \phi_0 + F A^y_{01}\, \phi_1, \label{eq:eig1} \\
E\, \phi_1 &= i F \frac{\partial \phi_1}{\partial k_y} + E_c\, \phi_1 + F A^y_{11}\, \phi_1 + F A^y_{10}\, \phi_0. \label{eq:eig2}
\end{align}
Here $\bm{\phi} = (\phi_0, \phi_1)^{\mathrm{T}}$ encodes the flat-band and dispersive-band components of the WS state, and $E$ is the WS eigenvalue. $E_c$ is given by Eq.~\eqref{eq:ec} and the flat band energy is zero. Eqs.~\eqref{eq:eig1} and ~\eqref{eq:eig2} resemble time-dependent Schr\"odinger equations, with $k_y$ playing the role of time. 

We focus on the $\Gamma$ valley ($|k_x|<\pi/2$); the X valley behaves similarly but with opposite chirality $\xi=-1$. The relevant Berry connections around $\Gamma$ can be approximated as (Sec.~I of the SM~\cite{sm})
\begin{align}
A^y_{00} = -A^y_{11} &\approx \sqrt{1-d^2}\,\frac{k_x}{k_x^2+k_y^2} + \pi\,\delta(k_y), \label{eq:berrys1} \\
A^y_{01} = A^y_{10} &\approx d\,\frac{k_x}{k_x^2+k_y^2}. \label{eq:berrys2}
\end{align}
The WS eigenproblem is solved with periodic boundary conditions in $k_y$,
\begin{align}
\bm{\phi} (k_x, -k_0) = \bm{\phi} (k_x, k_0),
\label{eq:peri}
\end{align}
where $2 k_0$ is the periodicity in $k$-space along the field direction~\cite{wsspec}; for the present two-band model, $k_0=\pi$. Below we solve Eqs.~\eqref{eq:eig1} and~\eqref{eq:eig2} analytically in two limits: (i) an isolated-flat-band regime away from the BCP, and (ii) a strongly coupled regime near the BCP where LZT becomes essential.

\begin{figure}
\centering
\includegraphics[width=8.6cm]{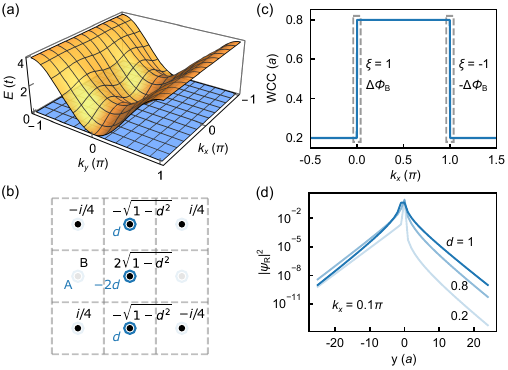}
\caption{(a) Band dispersion showing a flat band (blue) tangent to a dispersive band (orange) at $\Gamma$ $(0,0)$ and X $(\pi,0)$, with chirality (valley index) $\xi = \pm 1$, respectively. The band structure is independent of $d \in [0,1]$. (b) Compact localized state (CLS) of the flat band. Concentric blue and black dots denote two orbitals on each lattice site. Colored numbers indicate the wavefunction amplitudes of the CLS, which consists only of the orbitals shown in the corresponding dark color. Gray dashed lines mark unit cells. (c) Wannier charge center flow of the flat band versus $k_x$ for $d=0.8$. As $k_x$ crosses a singular BCP, the Berry phase jumps by $\xi \Delta \Phi_{\mathrm{B}}$. (d) Real-space WS wavefunction of the flat band versus $y$ (semi-log scale) at $k_x=0.1\pi$ for $d=1$, $0.8$, and $0.2$.}
\label{fig1}
\end{figure}

\emph{Isolated flat band limit.\textemdash}
For $k_x$ sufficiently far from the BCP and for weak fields, interband coupling is negligible and Eqs.~\eqref{eq:eig1} and~\eqref{eq:eig2} decouple~\cite{wsspec}. Solving Eq.~\eqref{eq:eig1} with the periodic boundary condition~\eqref{eq:peri} yields a WS ladder of the flat band (see Sec.~III of the SM~\cite{sm}):
\begin{align}
E = n \Delta + e F \gamma (k_x),
\label{eq:fbspec}
\end{align}
where $n$ is an integer, $\Delta = \hbar \omega_{\mathrm{B}}$ is the WS level spacing, and $\omega_{\mathrm{B}} = e F a / \hbar$ is the Bloch frequency. $\gamma (k_x)$ is the Wannier charge center (WCC) which is proportional to the Berry phase $\Phi_{\mathrm{B}} (k_x)$ accumulated across the BZ along $y$~\cite{rmpberry,polar1,polar2,rmppolar,zak,wilson3}:
\begin{align}
\gamma(k_x) = \frac{a}{2\pi}\,\Phi_{\mathrm{B}}(k_x) = \frac{1}{2k_0}\int_{-k_0}^{k_0} A^y_{00}\,dk_y \; (\mathrm{mod}\; a).
\label{eq:wcc}
\end{align}
For the present model,
\begin{align}
\Phi_{\mathrm{B}}(k_x) = \mathrm{sgn}(\sin k_x)\,\pi\sqrt{1-d^2}+\pi \; (\mathrm{mod}\; 2\pi).
\label{eq:phib}
\end{align}
The WCC gives the average position of the Wannier function, displaced from the atomic orbitals by the Berry phase and thus determined by $d$. As shown in Fig.~\ref{fig1}(c) for $d = 0.8$, $\gamma (k_x)$ exhibits an abrupt jump as $k_x$ crosses a BCP (gray dashed boxes). This reflects that each BCP carries a Berry phase $\xi \Delta \Phi_{\mathrm{B}}$, where $\Delta \Phi_{\mathrm{B}} = 2 \pi \sqrt{1-d^2}$ can take any value in $(0,2\pi)$ for $d\in(0,1)$~\cite{prbFBwav}. The closure of the WCC flow over the periodic BZ is ensured by paired BCPs of opposite chirality. Importantly, the abrupt jump at the BCP is merely an artifact of the isolated-band approximation: the interband Berry connections [Eq.~\eqref{eq:berrys2}] become non-negligible near the BCP. In the next section we show how the WS spectrum evolves through these regions once interband coupling is included.

Although the electric field perturbs the perfect destructive interference underlying the CLS, the WS eigenstates in this regime remain exponentially localized along the field direction. Fig.~\ref{fig1}(d) shows the real-space wavefunction weight $|\psi_{\mathrm{R}}(y)|^2$ at $k_x=0.1\pi$ for several values of $d$. The exponential localization implies suppressed dc transport away from the BCP, consistent with Kubo--Greenwood results in related systems~\cite{prb23fbcond}. 

\emph{LZT near the singular point.\textemdash}
Close to the singular BCP ($|k_x|$ small), the isolated-flat-band approximation fails because the interband coupling is no longer negligible. We therefore compute the WS spectrum by diagonalizing the finite-size real-space 1D Hamiltonian (Sec.~II A of the SM~\cite{sm}), which is equivalent to solving Eqs.~\eqref{eq:eig1} and~\eqref{eq:eig2} with a truncated Fourier expansion in $k_y$. Because $k_y$ evolves periodically over the BZ, the WS spectrum can also be viewed as a Floquet quasienergy with periodicity $\Delta=eFa$~\cite{prb24flo}.

Fig.~\ref{fig2}(a--d) shows the WS spectra for $F / t = 0.002$ and several values of $d$. For the trivial case $d = 0$, LZT is absent and the spectrum consists of two decoupled WS ladders originating from the dispersive and flat bands; the flat-band ladder is $k_x$ independent, as expected. For SFBs with $d>0$, quantum geometry enables LZT and hybridizes the ladders, producing anticrossing gaps. Far from the BCP, these gaps vanish as the interband coupling decreases, recovering Eq.~\eqref{eq:fbspec}. Approaching the BCP, the flat-band WS levels bend as they hybridize with dispersive-band levels (orange dashed guides), smoothing the unphysical jump seen in Fig.~\ref{fig1}(c). This bending resembles the anomalous spreading of Landau levels in SFBs~\cite{nature,25SFB}: in both cases, an external field drives repulsive interband coupling governed by quantum geometry. For $0<d<1$, the WS spectra are asymmetric under $k_x\to -k_x$, reflecting the chirality of the flat-band wavefunction. The field direction $\hat{\bm{y}}$ together with chirality $\xi\hat{\bm{z}}$ selects an orientation $\hat{\bm{y}}\times \xi\hat{\bm{z}}=\xi\hat{\bm{x}}$.

\begin{figure}
\centering
\includegraphics[width=8.6cm]{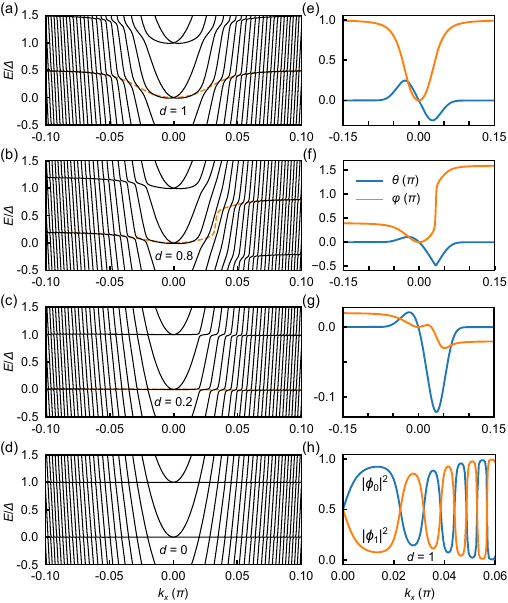}
\caption{(a--d) Numerical WS spectra (in units of $\Delta = e F a$) for $F/t = 0.002$ applied along $y$, plotted versus $k_x$ for several values of $d$. Orange dashed lines in (a--c) guide the eye for the flat-band WS levels and follow the dispersion set by $\varphi$. (e--g) Corresponding geometric phases $\theta$ (blue) and $\varphi$ (orange) for (a--c). (h) Band-resolved weights of the WS eigenstates for $d=1$.}
\label{fig2}
\end{figure}

To gain insights from quantum geometry, we analyze Eqs.~\eqref{eq:eig1} and~\eqref{eq:eig2} by treating intra- and interband Berry connections on an equal footing. For $k_x\to 0$, the Berry connections in Eqs.~\eqref{eq:berrys1} and~\eqref{eq:berrys2} are sharply peaked near $k_y = 0$ with width $\sim |k_x|$. We therefore approximate the hybridization as occurring only at $k_y=0$~\cite{wsspec}. The net effect of Berry connections across $k_y=0$ can then be encoded in a unitary rotation relating $\bm{\phi}$ at $k_y=0^-$ and $0^+$:
\begin{align}
\begin{pmatrix}
  \phi_0(0^+) \\
  \phi_1(0^+)
\end{pmatrix}
= U
\begin{pmatrix}
  \phi_0(0^-) \\
  \phi_1(0^-)
\end{pmatrix},
\label{eq:Uni1}
\end{align}
where
\begin{align}
U=
\begin{pmatrix}
  \cos\theta\,e^{i\varphi} & i\sin\theta \\[0.5em]
  i\sin\theta & \cos\theta\,e^{-i\varphi}
\end{pmatrix}.
\label{eq:umat}
\end{align}
The two geometric phases $(\theta,\varphi)$ fully characterize the LZT. The matrix $U$ follows from a path-ordered integral (Sec.~IV of the SM~\cite{sm}):
\begin{align}
U=-\mathcal{P}\exp\!\left[i\int_{-\infty}^{\infty}\frac{k_x}{k_x^2+k_y^2}\,\bm{w}\cdot\bm{\sigma}\,dk_y\right],
\label{eq:int}
\end{align}
where $\bm{w}=[d\cos(g_E),-d\sin(g_E),\sqrt{1-d^2}]$ is a unit vector and $\bm{\sigma}$ denotes Pauli matrices. We define
\begin{align}
g_E(k_x,k_y)=\frac{1}{F}\int_0^{k_y}E_c(k_x,k_y')\,dk_y',
\label{eq:ge}
\end{align}
analogous to a dynamical phase in time evolution. The integral is evaluated numerically (Sec.~V of the SM~\cite{sm}). Fig.~\ref{fig2}(e--g) plot $\theta$ and $\varphi$ versus $k_x$ alongside the WS spectra in Fig.~\ref{fig2}(a--c). For $d=0$, $\theta=\varphi=0$. Far from the BCP, $\theta\to 0$ and $\varphi\to\Phi_{\mathrm{B}}$, recovering the isolated-flat-band result. Close to the BCP, both phases contribute: $\theta$ quantifies interband tunneling, while $\varphi$ acts as a generalized Berry phase. Compared with conventional LZT, the tunneling probability is $P_{\mathrm{LZ}}=\sin^2\theta$, and $\varphi$ resembles the Stokes phase~\cite{pr10lz,pra97sto,pra15lz}. For SFBs, both phases are governed by $d$. Notably, $P_{\mathrm{LZ}}=0$ at a quadratic BCP~\cite{25fblz} and first increases and then decreases away from the BCP; this contrasts with linear crossings (e.g., graphene), where $P_{\mathrm{LZ}}=1$ at the BCP and decays exponentially away from it~\cite{prb12lz,pr10lz,pra97sto,pra15lz}. For $0<d<1$, $P_{\mathrm{LZ}}$ is larger along $+k_x$ due to chirality-induced asymmetry, leading to larger anticrossing gaps.

From Eqs.~\eqref{eq:int} and~\eqref{eq:ge}, $(\theta,\varphi)$ are functions of $F/t$ and $k_x$ and obey the scaling relations
\begin{align}
\begin{split}
\theta(\alpha^3 F/t, \alpha k_x) &= \theta(F/t, k_x), \\ 
\varphi(\alpha^3 F/t, \alpha k_x) &= \varphi(F/t, k_x),
\label{eq:sca}
\end{split}
\end{align}
with scaling parameter $\alpha$. Increasing $F$ expands the $k_x$ region where LZT is appreciable ($\theta\neq 0$). When this region becomes comparable to the full BZ ($F/t\sim 1$), the assumption that LZT occurs only near the BCP breaks down; thus, our analysis applies for $F/t\ll 1$.

Once $(\theta,\varphi)$ are known, incorporating the dynamical phase~\eqref{eq:ge} and imposing the periodic boundary condition~\eqref{eq:peri} yields the WS quantization condition:
\begin{align}
\cos [2 k_0 E / F - g_E (k_x, k_0)] = \cos \theta \cos [g_E (k_x, k_0) - \varphi].
\label{eq:spec}
\end{align}
Eq.~\eqref{eq:spec} reproduces Fig.~\ref{fig2}(a--d), showing explicitly that $\theta$ controls the gap size and $\varphi$ controls the energy shift. If we neglect gap opening by setting $\theta=0$, Eqs.~\eqref{eq:fbspec} and~\eqref{eq:wcc} apply with $\Phi_{\mathrm{B}}$ replaced by $\varphi$; the resulting dispersions are plotted as the orange dashed guides in Fig.~\ref{fig2}(a--c).

The band-resolved weights $|\phi_0|^2=\cos^2 \zeta$ and $|\phi_1|^2=\sin^2 \zeta$ of the WS eigenstates satisfy
\begin{align}
\sin 2 \zeta = \sin \theta / \sin [2 k_0 E / F - g_E (k_x, &k_0)],
\label{eq:wavf}
\end{align}
as illustrated in Fig.~\ref{fig2}(h) for $d=1$. Near each anticrossing point, strong interband hybridization mixes the localized flat-band state and the more extended dispersive-band state with real-space spread $\sqrt{\langle \bm{r}^2 \rangle - \langle \bm{r} \rangle^2}$ that scales as $t/F$~\cite{prl23bo,prb87ws}. Consequently, the real-space WS wavefunctions become delocalized (see Fig. S2 in the SM~\cite{sm}), leading to nontrivial flat band transport~\cite{prb25bo}.

\begin{figure}
\centering
\includegraphics[width=8.6cm]{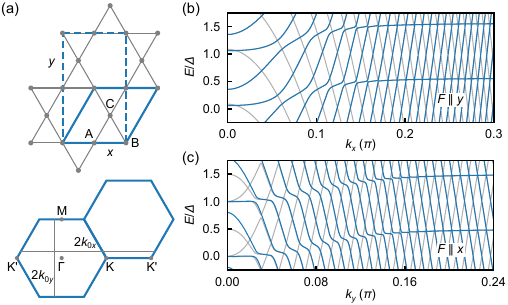}
\caption{(a) Upper: kagome lattice with $x$ and $y$ directions indicated. Solid and dashed blue boxes denote the primitive unit cell and the enlarged unit cell, respectively. Lower: BZ and high-symmetry points; gray lines indicate the $k$-space periodicities $2k_{0x}$ and $2k_{0y}$. (b),(c) WS spectra for $F / t = 0.01$ applied along $y$ and $x$, plotted versus $k_x$ and $k_y$ in units of $\Delta = \pi e F / k_{0y}$ and $\Delta = \pi e F / k_{0x}$, respectively. Gray lines: lattice calculation; blue lines: prediction from the geometric phases $(\theta,\varphi)$ for $d=1$.}
\label{fig3}
\end{figure}

\emph{Generalization to realistic models.\textemdash}
So far we have focused on a minimal two-band lattice model with tunable $d$, and show that the WS states are governed by two geometric phases $(\theta,\varphi)$. We now demonstrate that the same description applies straightforwardly to realistic SFB systems, using the kagome lattice as an example. We apply an electric field with $F / t = 0.01$ along $x$ or $y$ [Fig.~\ref{fig3}(a)], where $t$ is the nearest-neighbor hopping. The unit cell is enlarged (dashed blue box) to be compatible with the field directions. The WS spectra computed in the 1D real-space formalism are shown by gray lines in Fig.~\ref{fig3}(b) and (c) for $F \parallel y$ and $F \parallel x$, respectively. Each spectrum contains three WS ladders, two of which originate from the SFB with $d=1$. 

Using $(\theta,\varphi)$ for $d=1$ from Fig.~\ref{fig2}(e) together with the scaling relation~\eqref{eq:sca}, we obtain the WS spectra of the SFB from Eq.~\eqref{eq:spec} with $k_0=k_{0x}$ or $k_{0y}$ indicated in Fig.~\ref{fig3}(a) (see Sec.~VI of the SM~\cite{sm}). For $F \parallel y$, the blue curves in Fig.~\ref{fig3}(b) coincide with the two SFB ladders. For $F \parallel x$ [Fig.~\ref{fig3}(c)], the slight discrepancy at small $k_y$ arises from additional LZT between the two dispersive bands near the Dirac cones at the $\mathrm{K}$ and $\mathrm{K}'$ points, as indicated in Fig.~\ref{fig3}(a). 

In summary, we show that electric-field-driven LZT near a singular BCP is governed solely by the maximal quantum distance and encoded by two geometric phases, providing a quantum-geometric route to nontrivial flat-band transport. We provide a unified picture of the response of SFBs to electric and magnetic fields in disrupting the perfect destructive interference, shedding new light on the fundamental role of quantum geometry in flat-band physics.

\emph{Acknowledgments.\textemdash}
We thank Guangjie Li and Zheng Liu for helpful discussions. This work was fully supported by the DOE-BES (No. DE-FG02-04ER46148).

\bibliography{ref}

\newpage

\title{Supplementary Material for ``Quantum-geometry-enabled Landau--Zener tunneling in singular flat bands''}

\date{\today}

\maketitle

\onecolumngrid
\setcounter{section}{0}
\renewcommand{\thesection}{\Roman{section}}
\renewcommand{\thesubsection}{\Alph{subsection}}
\makeatletter
\renewcommand{\p@subsection}{\thesection\,}
\makeatother
\setcounter{secnumdepth}{2}

\setcounter{equation}{0}
\setcounter{figure}{0}

\renewcommand{\theequation}{S\arabic{equation}}
\renewcommand{\thefigure}{S\arabic{figure}}

\section{Details of the model Hamiltonian}\label{sec:model}
Following the general construction scheme of Ref.~\cite{fbcons}, we construct a minimal two-band lattice model hosting a singular flat band (SFB) with a tunable maximal quantum distance $d$:
\begin{equation}
H(\bm{k}) =
 t\begin{pmatrix}
  \bigl(1-\cos 2k_x\bigr)\bigl(1+\cos k_y\bigr)/4 + (1-d^2)\bigl(2-2\cos k_y\bigr)
  & d\sqrt{1-d^2}\,\bigl(2-2\cos k_y\bigr) - i d\sin k_x\sin k_y \\[0.5em]
  d\sqrt{1-d^2}\,\bigl(2-2\cos k_y\bigr) + i d\sin k_x\sin k_y
  & d^2\bigl(2-2\cos k_y\bigr)
 \end{pmatrix}.
\label{eq:laths}
\end{equation}
As shown in Fig.~1(a) of the main text, the spectrum consists of an exactly flat band (blue) and a dispersive band (orange) that touches it at two quadratic band-crossing points (BCPs) located at $\Gamma\,(0,0)$ and X$\,(\pi,0)$. Expanding Eq.~\eqref{eq:laths} around $\Gamma$ and X yields the effective continuum Hamiltonian
\begin{equation}
H_{\mathrm{SFB}} =
 t\begin{pmatrix}
  k_x^2 + (1-d^2)k_y^2 & d\sqrt{1-d^2}\,k_y^2 - i\,\xi d\,k_xk_y \\[0.5em]
  d\sqrt{1-d^2}\,k_y^2 + i\,\xi d\,k_xk_y & d^2k_y^2
 \end{pmatrix}.
\label{eq:sfbs}
\end{equation}
Here $\xi$ encodes the chirality of the SFB and can be regarded as a valley index: $\xi=+1$ for the $\Gamma$ point and $\xi=-1$ for the X point. The dispersive-band energy is
\begin{align}
E_c(\bm{k}) = t\,\bigl[9-7\cos k_y - \cos(2k_x)\bigl(1+\cos k_y\bigr)\bigr]/4.
\label{eq:ecs}
\end{align}
Expanding $E_c(\bm{k})$ around $\Gamma$, one finds $E_c(\bm{k})\approx t\,(k_x^2+k_y^2)$, corresponding to an isotropic parabolic dispersion.

The Bloch eigenvectors of the flat band and the dispersive band can be chosen as
\begin{align}
\bm{u}_0(\bm{k}) &= \frac{1}{c(\bm{k})}
\begin{pmatrix}
  -2d\,(1-\cos k_y)  \\[0.5em]
  2\sqrt{1-d^2}\,(1-\cos k_y) + i\sin k_x\sin k_y
\end{pmatrix},\label{eq:wflat1}\\
\bm{u}_1(\bm{k}) &= \frac{1}{c(\bm{k})}
\begin{pmatrix}
  2\sqrt{1-d^2}\,(1-\cos k_y) - i\sin k_x\sin k_y \\[0.5em]
  2d\,(1-\cos k_y)
\end{pmatrix},\label{eq:wflat2}
\end{align}
where
\begin{align}
 c(\bm{k}) = \sqrt{4(1-\cos k_y)^2 + \sin^2 k_x\sin^2 k_y}
\end{align}
normalizes the eigenvectors.

\begin{figure}
\centering
\includegraphics[width=8.6cm]{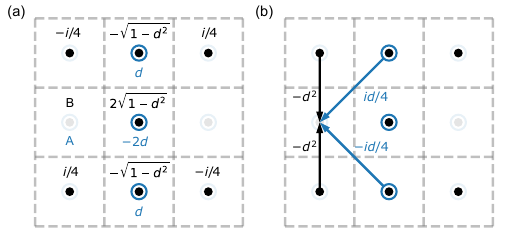}
\caption{(a) Compact localized state (CLS) of the flat band. Concentric blue and black dots denote two orbitals on each lattice site. Colored numbers indicate the wavefunction amplitudes of the CLS, which consists only of the orbitals shown in the corresponding dark color. Gray dashed lines mark unit cells. (b) A representative interference channel demonstrating perfect destructive interference. The arrows indicate hoppings to a $B$ orbital outside the CLS; the numbers and colors denote the hopping amplitudes and their originating orbitals, respectively.}
\label{S1}
\end{figure}

Fourier transforming the \emph{unnormalized} flat-band eigenvector in Eq.~\eqref{eq:wflat1} (i.e., setting $c(\bm{k})=1$) yields the corresponding compact localized state (CLS), shown in Fig.~\ref{S1}(a) [see also Fig.~1(b) in the main text]. One can explicitly verify that this CLS exhibits perfect destructive interference under the hopping amplitudes encoded in Eq.~\eqref{eq:laths}; a representative cancellation channel is illustrated in Fig.~\ref{S1}(b).

We also calculate the Berry connections $\bm{A}_{mn} (\bm{k})=i\langle u_m (\bm{k}) |\frac{\partial}{\partial \bm{k}}| u_n (\bm{k}) \rangle$, of which the $y$ components are
\begin{align}
A^y_{00}=-A^y_{11} &= \frac{2\sqrt{1-d^2}\,\sin k_x\,(1-\cos k_y)}{4(1-\cos k_y)^2+\sin^2 k_x\sin^2 k_y} + \pi\,\delta(k_y),\label{eq:berry1}\\
A^y_{01}=A^y_{10} &= \frac{2d\,\sin k_x\,(1-\cos k_y)}{4(1-\cos k_y)^2+\sin^2 k_x\sin^2 k_y}.
\label{eq:berry2}
\end{align}
The $\pi\,\delta(k_y)$ contribution originates from the discontinuity of the eigenvectors in Eqs.~\eqref{eq:wflat1} and~\eqref{eq:wflat2} across the line $k_y=0$. Expanding Eqs.~\eqref{eq:berry1} and~\eqref{eq:berry2} around $\Gamma$, we obtain
\begin{align}
A^y_{00}=-A^y_{11} &\approx \sqrt{1-d^2}\,\frac{k_x}{k_x^2+k_y^2} + \pi\,\delta(k_y),\label{eq:berrys1s}\\
A^y_{01}=A^y_{10} &\approx d\,\frac{k_x}{k_x^2+k_y^2}.
\label{eq:berrys2s}
\end{align}

\section{Hamiltonian under an electric field}\label{sec:efield}
Here we derive the Hamiltonian of a generic lattice model in a static, uniform electric field $\bm{F}$~\cite{prl23bo,wsspec}. We first formulate the problem in real space and then derive the corresponding eigenvalue equations in momentum space.

\subsection{Real-space formulation}\label{sec:hreal}
We start from a general $M$-band tight-binding model. In real space, each unit cell contains $M$ atomic orbitals $\bigl|\bm{R},i\bigr\rangle$ located at $\bm{R}+\bm{\tau}_i$, where $\bm{R}=m\bm{a}_1+n\bm{a}_2$ labels the unit cell ($m,n\in\mathbb{Z}$) and $\bm{\tau}_i$ is the intra-cell displacement of orbital $i$.

For the minimal two-band model, $\bm{a}_1=(a,0)$, $\bm{a}_2=(0,a)$, and $\bm{\tau}_i=\bm{0}$ for $i=A,B$ [see Fig.~\ref{S1}(a)], where $a$ is the lattice constant (set to $a=1$ below unless otherwise stated). The real-space hopping parameters are $t_{\Delta\bm{R},ij}=\bigl\langle \bm{R},i\big|H\big|\bm{R}+\Delta\bm{R},j\bigr\rangle$, and translational symmetry implies $t_{\bm{R},ij}=\langle \bm{0},i|H|\bm{R},j\rangle$.

With the Fourier transforms
\begin{align}
\bigl|\bm{k},i\bigr\rangle &= \frac{1}{\sqrt{N}}\sum_{\bm{R}} e^{i\bm{k}\cdot(\bm{R}+\bm{\tau}_i)}\,\bigl|\bm{R},i\bigr\rangle,\label{eq:phik}\\
\bigl|\bm{R},i\bigr\rangle &= \frac{1}{\sqrt{N}}\sum_{\bm{k}} e^{-i\bm{k}\cdot(\bm{R}+\bm{\tau}_i)}\,\bigl|\bm{k},i\bigr\rangle,\label{eq:wr}
\end{align}
we relate the momentum-space Hamiltonian matrix elements $[H(\bm{k})]_{ij}=\langle \bm{k},i|H| \bm{k},j\rangle$ to the real-space hoppings via
\begin{align}
[H(\bm{k})]_{ij} &= \sum_{\bm{R}} e^{i\bm{k}\cdot(\bm{R}+\bm{\tau}_j-\bm{\tau}_i)}\,t_{\bm{R},ij},\label{eq:hk}\\
 t_{\bm{R},ij} &= \sum_{\bm{k}} e^{-i\bm{k}\cdot(\bm{R}+\bm{\tau}_j-\bm{\tau}_i)}\,[H(\bm{k})]_{ij}.
\label{eq:hr}
\end{align}

A uniform electric field adds the scalar-potential term
\begin{align}
H' = e\,\bm{F}\cdot\bm{r}.
\label{eq:hp}
\end{align}
For compactness we set $e=1$ in intermediate steps; it can be restored by the replacement $\bm{F}\to e\bm{F}$. In the basis $\bigl|\bm{R},i\bigr\rangle$, the corresponding matrix elements read
\begin{align}
\bigl\langle\bm{R},i\big|\bm{F}\cdot\bm{r}\big|\bm{R}',j\bigr\rangle = \bm{F}\cdot(\bm{R}+\bm{\tau}_i)\,\delta_{ij}\,\delta_{\bm{R}\bm{R}'},
\label{eq:hpr}
\end{align}
which shows that the electric field produces a position-dependent onsite potential along the field direction.

For the minimal two-band model we take $\bm{F}=F\,\hat{\bm{y}}$. Translational symmetry along $x$ remains intact, so $k_x$ is a good quantum number. We therefore block-diagonalize the Hamiltonian at fixed $k_x$ and reduce the problem to a one-dimensional (1D) chain along $y$ using the basis
\begin{align}
\bigl|k_x,y,i\bigr\rangle &= \frac{1}{\sqrt{N_x}}\sum_x e^{ik_x(x+\tau_{i,x})}\,\bigl|\bm{R},i\bigr\rangle
\label{eq:wkr}\\
&= \frac{1}{\sqrt{N_y}}\sum_{k_y} e^{-ik_y(y+\tau_{i,y})}\,\bigl|\bm{k},i\bigr\rangle.
\label{eq:wkr2}
\end{align}
At a given $k_x$, we construct a sufficiently long 1D cluster along $y$ with onsite potential $V_{n,i}=F(na+\tau_{i,y})$ and diagonalize the resulting real-space Hamiltonian. The Wannier--Stark (WS) spectra shown in Fig.~2(a--d) of the main text are obtained in this way; the hopping parameters are generated from Eq.~\eqref{eq:laths} using the inverse Fourier transform in Eq.~\eqref{eq:hr}. We compute the WS spectra for the kagome lattice in Fig.~3(b) and (c) of the main text similarly, choosing the enlarged unit cell as the dashed blue box in the upper panel of Fig.~3(a) so that the basis vectors align with the $x$ or $y$ direction.

\subsection{Momentum-space formulation}\label{sec:hkspace}
Under the orbital basis $\bigl|\bm{k},i\bigr\rangle$ defined by Eq.~\eqref{eq:phik}, the field-free Hamiltonian is block-diagonal in $\bm{k}$ and given by Eq.~\eqref{eq:laths}. The electric-field term has matrix elements
\begin{align}
\langle\bm{k},i|\bm{F}\cdot\bm{r}|\bm{k},j\rangle = i\,\bm{F}\cdot\nabla_{\bm{k}}\,\delta_{ij}.
\label{eq:hpk}
\end{align}
It is convenient to switch to the band basis
\begin{align}
\bigl|u_m(\bm{k})\bigr\rangle = \sum_i u_{m,i}(\bm{k})\,\bigl|\bm{k},i\bigr\rangle,
\label{eq:wfpsi}
\end{align}
where $m\in\{0,1\}$, $i\in\{A,B\}$, and $u_{m,i}(\bm{k})$ are given in Eqs.~\eqref{eq:wflat1} and~\eqref{eq:wflat2}. In this basis the field-free Hamiltonian is diagonal, with eigenvalues $0$ and $E_c$ [Eq.~\eqref{eq:ecs}]. The electric-field matrix elements read
\begin{align}
\bigl\langle u_m(\bm{k})\big|\bm{F}\cdot\bm{r}\big|u_n(\bm{k})\bigr\rangle = i\,\bm{F}\cdot\nabla_{\bm{k}}\,\delta_{mn} + \bm{F}\cdot\bm{A}_{mn}(\bm{k}),
\label{eq:hpk2}
\end{align}
with $\bm{A}_{mn}$ the Berry connections [Eqs.~\eqref{eq:berry1} and~\eqref{eq:berry2}].

We expand an eigenstate in the band basis as
\begin{align}
|\Psi\rangle = \frac{1}{\sqrt{N}}\sum_{\bm{k}}\Bigl[\phi_0(\bm{k})\,|u_0(\bm{k})\rangle + \phi_1(\bm{k})\,|u_1(\bm{k})\rangle\Bigr],
\label{eq:psi}
\end{align}
with the normalization condition $|\phi_0|^2 + |\phi_1|^2 = 1$. Using Eqs.~\eqref{eq:ecs} and~\eqref{eq:hpk2}, the eigenvalue equation $(H+H')|\Psi\rangle=E|\Psi\rangle$ becomes
\begin{align}
E\,\phi_0 &= iF\,\frac{\partial\phi_0}{\partial k_y} + F A^y_{00}\,\phi_0 + F A^y_{01}\,\phi_1,\label{eq:eig1s}\\
E\,\phi_1 &= iF\,\frac{\partial\phi_1}{\partial k_y} + E_c\,\phi_1 + F A^y_{11}\,\phi_1 + F A^y_{10}\,\phi_0,
\label{eq:eig2s}
\end{align}
where we have taken $\bm{F}=F\,\hat{\bm{y}}$. At fixed $k_x$, the eigenvector $\bm{\phi}=(\phi_0,\phi_1)^{\mathrm{T}}$ is a function of $k_y$ with period $2k_0$, set by the $k$-space periodicity along the field direction. For the minimal two-band model, $k_0=\pi$. For the kagome lattice, $2k_0$ depends on the field direction and is indicated in the lower panel of Fig.~3(a) of the main text. The WS spectrum $E(k_x)$ is obtained by solving Eqs.~\eqref{eq:eig1s} and~\eqref{eq:eig2s} subject to the periodic boundary condition
\begin{align}
\bm{\phi}(k_x,-k_0)=\bm{\phi}(k_x,k_0).
\label{eq:peris}
\end{align}

\section{Isolated-flat-band limit}\label{sec:iso}
Away from the BCP, the interband coupling can be neglected to leading order. Dropping the interband term $A^y_{01}$ in Eq.~\eqref{eq:eig1s}, we obtain a closed equation for the flat band. The eigenstate in Eq.~\eqref{eq:psi} then has $\phi_1(\bm{k})=0$ and
\begin{align}
\phi_0(\bm{k}) = \exp\left[-\frac{iE}{F}k_y + i\int_{-k_0}^{k_y} A^y_{00}(k_x,k_y')\,dk_y'\right],
\label{eq:phi0k}
\end{align}
where $k_0=\pi$ and
\begin{align}
\int_{-k_0}^{k_y} A^y_{00}(k_x,k_y')\,dk_y'
= \sqrt{1-d^2}\,\arctan\!\left(\frac{2\tan(k_y'/2)}{\sin k_x}\right)\biggr|_{-k_0}^{k_y}
+ \frac{\pi}{2}\bigl[1+\mathrm{sgn}(k_y)\bigr].
\label{eq:intg}
\end{align}
For the special case $k_y=k_0$, Eq.~\eqref{eq:intg} reduces to the Berry phase $\Phi_{\mathrm{B}}(k_x)$,
\begin{align}
\Phi_{\mathrm{B}}(k_x) = \mathrm{sgn}(\sin k_x)\,\pi\sqrt{1-d^2} + \pi.
\label{eq:phibs}
\end{align}
Imposing the periodicity condition $\phi_0(k_x,-k_0)=\phi_0(k_x,k_0)$ [Eq.~\eqref{eq:peris}] yields the WS energies
\begin{align}
E = \left\{n + \bigl[\mathrm{sgn}(\sin k_x)\sqrt{1-d^2}+1\bigr]/2\right\}\,eFa,
\label{eq:ene}
\end{align}
where $n\in\mathbb{Z}$. This result is equivalent to Eq.~(9) in the main text expressed in terms of the Wannier charge center.

We now evaluate the real-space profile of the eigenstate and show that it is exponentially localized. At fixed $k_x$, Eq.~\eqref{eq:psi} gives
\begin{align}
|\Psi_{k_x}\rangle = \frac{1}{\sqrt{N_y}}\sum_{k_y}\Bigl[\phi_0(\bm{k})\,|u_0(\bm{k})\rangle + \phi_1(\bm{k})\,|u_1(\bm{k})\rangle\Bigr].
\label{eq:psikx}
\end{align}
The corresponding one-dimensional real-space wavefunction $\psi_{\mathrm{R},i}(y)=\langle k_x,y,i|\Psi_{k_x}\rangle$ follows from Eqs.~\eqref{eq:wkr2} and~\eqref{eq:wfpsi}:
\begin{align}
\psi_{\mathrm{R},i}(y) = \frac{1}{N_y}\sum_{k_y} e^{ik_y(y+\tau_{i,y})}\,\Bigl[\phi_0(\bm{k})u_{0,i}(\bm{k}) + \phi_1(\bm{k})u_{1,i}(\bm{k})\Bigr].
\label{eq:ft1d}
\end{align}
Here $\phi_0(\bm{k})$ is given by Eq.~\eqref{eq:phi0k}, $\phi_1(\bm{k})=0$, and $\tau_{i,y}=0$ for the minimal two-band model. Defining $|\psi_{\mathrm{R}}|^2=\sum_i|\psi_{\mathrm{R},i}|^2$, we obtain the exponentially localized profile shown in Fig.~1(d) of the main text for $k_x=0.1\pi$. We have also verified that Eq.~\eqref{eq:ft1d} agrees with the wavefunction obtained by direct diagonalization of the real-space Hamiltonian in Sec.~\ref{sec:hreal} in the weak-field (uncoupled) regime.

\section{Solution for two coupled bands near the singular point}\label{sec:coup}
We now solve the coupled eigenvalue equations~\eqref{eq:eig1s} and~\eqref{eq:eig2s} in the vicinity of a singular BCP. Define
\begin{align}
 g_E(k_x,k_y)=\frac{1}{F}\int_0^{k_y} E_c(k_x,k_y')\,dk_y'.
\label{eq:ges}
\end{align}
For $k_x,k_y\to 0$, one has $g_E\approx t\,(k_y^3/3+k_x^2k_y)/F$. Introduce the transformations
\begin{align}
\tilde{\phi}_0 = \phi_0\,e^{iEk_y/F},\qquad
\tilde{\phi}_1 = \phi_1\,e^{iEk_y/F-ig_E}.
\label{eq:tilde}
\end{align}
Then Eqs.~\eqref{eq:eig1s} and~\eqref{eq:eig2s} become
\begin{align}
 i\frac{\partial\tilde{\phi}_0}{\partial k_y} + A^y_{00}\,\tilde{\phi}_0 + A^y_{01}e^{ig_E}\,\tilde{\phi}_1 &= 0,
\label{eq:eig21}\\
 i\frac{\partial\tilde{\phi}_1}{\partial k_y} + A^y_{11}\,\tilde{\phi}_1 + A^y_{10}e^{-ig_E}\,\tilde{\phi}_0 &= 0.
\label{eq:eig22}
\end{align}
Using the Berry connections expanded around $\Gamma$ [Eqs.~\eqref{eq:berrys1s} and~\eqref{eq:berrys2s}], the coupled equations can be written compactly as
\begin{align}
 i\frac{\partial}{\partial k_y}\begin{pmatrix}\tilde{\phi}_0\\\tilde{\phi}_1\end{pmatrix}
 + \left[\frac{k_x}{k_x^2+k_y^2}\,\bm{w}\cdot\bm{\sigma} + \pi\,\delta(k_y)\,\sigma_z\right]\begin{pmatrix}\tilde{\phi}_0\\\tilde{\phi}_1\end{pmatrix}=0,
\label{eq:pde0}
\end{align}
where $\bm{\sigma}=(\sigma_x,\sigma_y,\sigma_z)$ are Pauli matrices and
\begin{align}
\bm{w}=\bigl[d\cos(g_E),\,-d\sin(g_E),\,\sqrt{1-d^2}\bigr]
\end{align}
is a unit vector. The prefactor $k_x/(k_x^2+k_y^2)$ is sharply peaked near $k_y=0$, so we approximate the solution by two $k_y$-independent vectors, $\tilde{\bm{\phi}}(+)$ for $k_y>0$ and $\tilde{\bm{\phi}}(-)$ for $k_y<0$, related by
\begin{align}
\begin{pmatrix}\tilde{\phi}_0(+)\\\tilde{\phi}_1(+)\end{pmatrix}
= U\begin{pmatrix}\tilde{\phi}_0(-)\\\tilde{\phi}_1(-)\end{pmatrix}.
\label{eq:Uni1s}
\end{align}
The unitary rotation matrix $U$ is obtained from the path-ordered exponential
\begin{align}
U = -\mathcal{P}\exp\left[i\int_{-\infty}^{\infty}\frac{k_x}{k_x^2+k_y^2}\,\bm{w}\cdot\bm{\sigma}\,dk_y\right].
\label{eq:ints}
\end{align}
This integral is evaluated numerically in Sec.~\ref{sec:nume}. One can parametrize
\begin{align}
U = \begin{pmatrix}
\cos\theta\,e^{i\varphi} & i\sin\theta\\[0.3em]
 i\sin\theta & \cos\theta\,e^{-i\varphi}
\end{pmatrix},
\label{eq:umats}
\end{align}
where $\theta$ and $\varphi$ depend on $F/t$, $k_x$, and $d$.

Consider the scaling $\bm{k}\to\alpha\bm{k}$ in Eq.~\eqref{eq:ints}. The factor $k_x\,dk_y/(k_x^2+k_y^2)$ is invariant, while invariance of $g_E$ near the quadratic BCP requires $F/t\to\alpha^3F/t$ because $g_E$ scales cubically with $\bm{k}$. Therefore,
\begin{align}
\theta(\alpha^3F/t,\alpha k_x)=\theta(F/t,k_x),
\qquad
\varphi(\alpha^3F/t,\alpha k_x)=\varphi(F/t,k_x).
\label{eq:scas}
\end{align}

Once $\theta$ and $\varphi$ are known, the WS spectrum follows from the periodic boundary condition in Eq.~\eqref{eq:peris}. We take
\begin{align}
\begin{pmatrix}\phi_0(-k_0)\\\phi_1(-k_0)\end{pmatrix}
=\begin{pmatrix}\phi_0(k_0)\\\phi_1(k_0)\end{pmatrix}
=\begin{pmatrix}\cos\zeta\\\sin\zeta\end{pmatrix},
\label{eq:bound}
\end{align}
where $k_0=\pi$ for the minimal two-band model and we choose a real, normalized eigenvector parameterized by $\zeta$.
Using Eq.~\eqref{eq:tilde}, we have
\begin{align}
\begin{pmatrix}\tilde{\phi}_0(+)\\\tilde{\phi}_1(+)\end{pmatrix}
=\begin{pmatrix}e^{i\theta_0}\cos\zeta\\ e^{i\theta_1}\sin\zeta\end{pmatrix},
\qquad
\begin{pmatrix}\tilde{\phi}_0(-)\\\tilde{\phi}_1(-)\end{pmatrix}
=\begin{pmatrix}e^{-i\theta_0}\cos\zeta\\ e^{-i\theta_1}\sin\zeta\end{pmatrix},
\label{eq:twowf}
\end{align}
where $\theta_0=Ek_0/F$ and $\theta_1=Ek_0/F-g_E(k_x,k_0)$. Combining Eqs.~\eqref{eq:Uni1s}, \eqref{eq:umats}, and \eqref{eq:twowf}, the solvability condition for $\zeta$ yields
\begin{align}
\cos [2 k_0 E / F - g_E (k_x, k_0)] = \cos \theta \cos [g_E (k_x, k_0) - \varphi],
\label{eq:specs}
\end{align}
which determines the WS energies. The mixing angle $\zeta$ follows from
\begin{align}
\sin 2 \zeta = \sin \theta / \sin [2 k_0 E / F - g_E (k_x, &k_0)],
\label{eq:wavfs}
\end{align}
quantifying the hybridization between the flat and dispersive bands.

Finally, based on Eqs.~\eqref{eq:tilde} and~\eqref{eq:twowf}, $\bm{\phi}(k_y)$ is solved as:
\begin{align}
\begin{pmatrix}\phi_0(k_y)\\\phi_1(k_y)\end{pmatrix}
=\begin{pmatrix}
 e^{-iEk_y/F+i\theta_0\,\mathrm{sgn}(k_y)}\cos\zeta\\
 e^{-iEk_y/F+ig_E+i\theta_1\,\mathrm{sgn}(k_y)}\sin\zeta
\end{pmatrix}.
\label{eq:phi01}
\end{align}
Thus $|\phi_0|^2=\cos^2\zeta$ and $|\phi_1|^2=\sin^2\zeta$ are independent of $k_y$; for $d=1$ they are plotted in Fig.~2(h) of the main text using the $\theta$ and $\varphi$ shown in Fig.~2(e).

\begin{figure}[H]
\centering
\includegraphics[width=17.0cm]{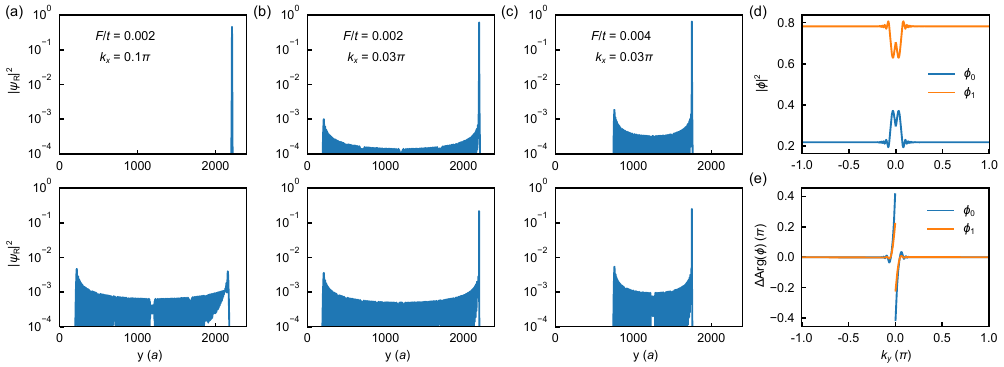}
\caption{(a) Two representative real-space wavefunctions with adjacent energies, plotted as a function of $y$ on a semi-log scale, for $d=1$, $F/t=0.002$, and $k_x=0.1\pi$. (b) Same as (a) but with $k_x=0.03\pi$. (c) Same as (b) but with $F/t=0.004$. (d) Band-resolved weights $|\phi_{0,1}|^2$ as a function of $k_y$, corresponding to the lower panel of (b). (e) Phase difference $\Delta\mathrm{Arg}(\phi_{0,1})$ between the exact numerical solution and the analytical prediction in Eq.~\eqref{eq:phi01}, corresponding to (d).}
\label{S2}
\end{figure}

On the other hand, following the procedure described at the end of Sec.~\ref{sec:hreal}, the WS spectra and the corresponding real-space wavefunctions $\psi_{\mathrm{R},i}(y)$ can be obtained numerically at fixed $k_x$. The resulting WS spectra shown in Fig.~2(a--d) of the main text agree with the analytical prediction in Eq.~\eqref{eq:specs}. For the real-space wavefunctions, we again define $|\psi_{\mathrm{R}}|^2=\sum_i|\psi_{\mathrm{R},i}|^2$ and plot it as a function of $y$ in Fig.~\ref{S2}(a--c) for several $k_x$ and $F/t$ values (with $d=1$). Translating $\psi_{\mathrm{R},i}(y)$ by one unit cell along $y$ yields another eigenstate whose energy is shifted by $\Delta=eFa$; therefore, only two representative wavefunctions with adjacent energies are shown.

In the isolated-flat-band limit ($F/t=0.002$ and $k_x=0.1\pi$) [Fig.~\ref{S2}(a)], we find an exponentially localized wavefunction associated with the flat band together with a much more extended wavefunction originating from the dispersive band. Closer to the BCP at $k_x=0.03\pi$ [Fig.~\ref{S2}(b)], strong interband hybridization mixes the localized flat-band state with the more extended dispersive-band state, leading to a pronounced delocalization of the SFB wavefunction. Increasing the field strength to $F/t=0.004$ [Fig.~\ref{S2}(c)], the dispersive-band state becomes more loclaized, with a real-space spread $\sqrt{\langle\bm{r}^2\rangle-\langle\bm{r}\rangle^2}$ that scales as $t/F$~\cite{prl23bo,prb87ws}.

The exact solutions for $\phi_0(k_y)$ and $\phi_1(k_y)$ can be extracted from $\psi_{\mathrm{R},i}(y)$ via the inverse Fourier transform of Eq.~\eqref{eq:ft1d}:
\begin{align}
\phi_0(\bm{k})u_{0,i}(\bm{k})+\phi_1(\bm{k})u_{1,i}(\bm{k}) = \sum_y e^{-ik_y(y+\tau_{i,y})}\,\psi_{\mathrm{R},i}(y).
\label{eq:ftinv}
\end{align}
Comparing with Eq.~\eqref{eq:phi01}, we find that away from $k_y=0$, $|\phi_0|^2$ and $|\phi_1|^2$ approach constants [Fig.~\ref{S2}(d)], in agreement with $\cos^2\zeta$ and $\sin^2\zeta$, respectively. The phases of $\phi_0$ and $\phi_1$ also match the analytical prediction in Eq.~\eqref{eq:phi01} [Fig.~\ref{S2}(e)], validating the characterization of LZT in terms of $\theta$ and $\varphi$.

\section{Numerical recipe to calculate $\theta$ and $\varphi$}\label{sec:nume}
This section details how we compute $\theta$ and $\varphi$ from the path-ordered exponential in Eq.~\eqref{eq:ints}. Because the Pauli matrices $\sigma_x$, $\sigma_y$, and $\sigma_z$ do not commute, we use a Trotter discretization with finite $\Delta k_y$~\cite{numlz}. We truncate the integral at $\pm k_{y,\mathrm{max}}$ and divide the interval into $N$ steps so that $N\Delta k_y=2k_{y,\mathrm{max}}$. Direct evaluation of Eq.~\eqref{eq:ints} converges slowly with respect to both $k_{y,\mathrm{max}}$ and $N$.

To improve convergence, we first integrate the $\sigma_z$ component and define
\begin{align}
\hat{\phi}_0=\tilde{\phi}_0 e^{-ig_A},\qquad
\hat{\phi}_1=\tilde{\phi}_1 e^{ig_A},
\label{eq:phihat}
\end{align}
where
\begin{align}
 g_A(k_x,k_y)=\int_0^{k_y}\sqrt{1-d^2}\,\frac{k_x}{k_x^2+k_y'^2}\,dk_y'
=\sqrt{1-d^2}\,\arctan(k_y/k_x).
\label{eq:ga}
\end{align}
Equation~\eqref{eq:pde0} then becomes
\begin{align}
 i\frac{\partial}{\partial k_y}\begin{pmatrix}\hat{\phi}_0\\\hat{\phi}_1\end{pmatrix}
 + \left[\frac{k_x}{k_x^2+k_y^2}\,\tilde{\bm{w}}\cdot\bm{\sigma} + \pi\,\delta(k_y)\,\sigma_z\right]\begin{pmatrix}\hat{\phi}_0\\\hat{\phi}_1\end{pmatrix}=0,
\label{eq:pde2}
\end{align}
with
\begin{align}
\tilde{\bm{w}}=\bigl[d\cos(g_E-2g_A),\,-d\sin(g_E-2g_A),\,0\bigr].
\end{align}
As before, $\hat{\bm{\phi}}(+)$ and $\hat{\bm{\phi}}(-)$ are related by
\begin{align}
\tilde{U}=-\mathcal{P}\exp\left[i\int_{-\infty}^{\infty}\frac{k_x}{k_x^2+k_y^2}\,\tilde{\bm{w}}\cdot\bm{\sigma}\,dk_y\right].
\label{eq:int2}
\end{align}
Using Eq.~\eqref{eq:phihat}, one finds
\begin{align}
U_{00}=\tilde{U}_{00}e^{i\pi\sqrt{1-d^2}\,\mathrm{sgn}(k_x)},\quad
U_{11}=\tilde{U}_{11}e^{-i\pi\sqrt{1-d^2}\,\mathrm{sgn}(k_x)},\quad
U_{01}=\tilde{U}_{01},\quad
U_{10}=\tilde{U}_{10}.
\end{align}

We evaluate Eq.~\eqref{eq:int2} numerically using the Trotter product
\begin{align}
\tilde{U}=\tilde{U}(-k_{y,\mathrm{max}},k_{y,\mathrm{max}})=\prod_{i=0}^{N-1}\tilde{U}\bigl(-k_{y,\mathrm{max}}+i\Delta k_y,\,-k_{y,\mathrm{max}}+(i+1)\Delta k_y\bigr),
\label{eq:prod}
\end{align}
where
\begin{align}
\begin{split}
\tilde{U}(k_y,k_y+\Delta k_y)
&=\exp\left[i\,\frac{k_x\Delta k_y}{k_x^2+k_y^2}\,\tilde{\bm{w}}\cdot\bm{\sigma}\right]\\
&=\cos\!\left(\frac{dk_x\Delta k_y}{k_x^2+k_y^2}\right)I
+i\sin\!\left(\frac{dk_x\Delta k_y}{k_x^2+k_y^2}\right)
\Bigl[\cos(g_E-2g_A)\,\sigma_x-\sin(g_E-2g_A)\,\sigma_y\Bigr].
\label{eq:prod2}
\end{split}
\end{align}
We take $k_{y,\mathrm{max}}=\pi$ and $N=10^4$ to compute the discretized product in Eq.~\eqref{eq:prod2}. The resulting $\theta$ and $\varphi$ are plotted in Fig.~2(e--g) of the main text.

\section{WS spectra from the SFB in the kagome lattice}\label{sec:kag}
For the kagome lattice with nearest-neighbor hopping amplitude $t_0$, the tight-binding Hamiltonian is
\begin{align}
H_{\mathrm{kag}}(\bm{k})=t_0\begin{pmatrix}
 0 & 2\cos(k_x/2) & 2\cos\!\left(k_x/4+\sqrt{3}k_y/4\right)\\[0.5em]
 2\cos(k_x/2) & 0 & 2\cos\!\left(-k_x/4+\sqrt{3}k_y/4\right)\\[0.5em]
 2\cos\!\left(k_x/4+\sqrt{3}k_y/4\right) & 2\cos\!\left(-k_x/4+\sqrt{3}k_y/4\right) & 0
\end{pmatrix},
\label{eq:hkag}
\end{align}
written in the orbital basis $|\bm{k},i\rangle$ with $i\in\{A,B,C\}$. The orbital positions within the primitive unit cell are $\bm{\tau}_A=(1/2,0)$, $\bm{\tau}_B=(1,0)$, and $\bm{\tau}_C=(3/4,\sqrt{3}/4)$, and we set the lattice constant $a=1$ [upper panel of Fig.~3(a) in the main text]. The band dispersions are $E_0=-2t_0$ and
\begin{align}
E_{1,2}=t_0\left[1\mp\sqrt{3+2\cos k_x+4\cos(k_x/2)\cos(\sqrt{3}k_y/2)}\right].
\label{eq:ekag}
\end{align}
Thus $E_c=E_1-E_0$ in Eq.~\eqref{eq:ges}. The effective continuum Hamiltonian of the SFB is given by Eq.~\eqref{eq:sfbs} with $d=1$ and $t=t_0/4$.

Under an electric field $F_0$, the corresponding dimensionless field strength is $F_0/t=4F_0/t_0$. Using the scaling relation in Eq.~\eqref{eq:scas}, we introduce the scaling parameter
\begin{align*}
\alpha=\sqrt[3]{\frac{4F_0/t_0}{F/t}}.
\end{align*}

We compute the Berry phase (Zak phase~\cite{zak}) of the flat band and the dispersive band tangent to it along an open path of length $2k_{0x}=4\pi$ (field along $x$) or $2k_{0y}=4\pi/\sqrt{3}$ (field along $y$) [lower panel of Fig.~3(a) in the main text]. A convenient discrete expression is~\cite{jpcm00zak,prb21zak}
\begin{align}
\Phi_{\mathrm{B}}=\mathrm{Im}\log\prod_{i=0}^{N-1}\langle u_m(\bm{k}_{i+1})|u_m(\bm{k}_i)\rangle
+\mathrm{Im}\log\langle u_m(\bm{k}_0)|e^{i\bm{G}\cdot\bm{r}}|u_m(\bm{k}_N)\rangle,
\label{eq:zak}
\end{align}
where $m\in\{0,1\}$ labels the band, $|u_m(\bm{k})\rangle$ is an eigenstate of Eq.~\eqref{eq:hkag}, and $\bm{k}_i$ are discretized points along the chosen path. The reciprocal lattice vector is $\bm{G}=(2k_{0x},0)$ for the $x$ path and $\bm{G}=(0,2k_{0y})$ for the $y$ path. In the orbital basis, the ``closure'' operator is
\begin{align}
 e^{i\bm{G}\cdot\bm{r}}=\mathrm{Diag}\bigl(e^{i\bm{G}\cdot\bm{\tau}_A},\,e^{i\bm{G}\cdot\bm{\tau}_B},\,e^{i\bm{G}\cdot\bm{\tau}_C}\bigr).
\end{align}

Along $x$, the Berry phases of both bands are $\pi$, identical to the minimal two-band model with $d=1$ discussed in the main text. Therefore, the scaled $\theta$ and $\varphi$ from Fig.~2(e), together with Eqs.~\eqref{eq:ges} and~\eqref{eq:specs}, can be applied directly to obtain the WS spectra of the SFB, shown by the blue lines in Fig.~3(c) of the main text. Note that the WS spectra of the SFB are shifted by $-2t_0$ for comparison with the lattice-model results, as the flat-band energy in the kagome model is nonzero.

Along $y$, however, the flat band has Berry phase $\pi$ while the dispersive band has Berry phase $0$; the total Berry phase within the two-band SFB subspace is not compensated because of the presence of the third band. Relative to the minimal two-band model, this corresponds to an additional $\pi$ Berry phase associated only with the dispersive band. We account for this by shifting $g_E\to g_E+\pi/2$ in Eq.~\eqref{eq:ges}, which yields the WS spectra shown by the blue lines in Fig.~3(b) of the main text. 

\end{document}